\documentclass[prl, showpacs, twocolumn, floatfix]{revtex4}
\usepackage{amsmath}
\usepackage{graphicx}
\usepackage{amsmath, amsfonts, amssymb, bm}
\begin{document}
\title{Single-Cycle Gap Soliton in a Subwavelength Structure}

\author{Xiao-Tao \surname{Xie}}
\email{xtxie@nwu.edu.cn}

\author{Mihai A. \surname{Macovei}}
\email{mihai.macovei@mpi-hd.mpg.de}

\affiliation{Max-Planck-Institut f\"{u}r Kernphysik, Saupfercheckweg
1, D-69117 Heidelberg, Germany}

\date{\today}
\begin{abstract}
We demonstrate that a single sub-cycle optical pulse can be generated 
when a pulse with a few optical cycles penetrates through resonant
two-level dense media with a subwavelength structure. The
single-cycle gap soliton phenomenon in the full Maxwell-Bloch
equations without the frame of slowly varying envelope and rotating
wave approximations is observed. Our study shows that the
subwavelength structure can be used to suppress the frequency shift
caused by intrapulse four-wave mixing in continuous media and
supports the formation of single-cycle gap solitons even in the case
when the structure period breaks the Bragg condition. This suggests a
way toward shortening high-intensity laser fields to few- and even
single-cycle pulse durations.
\end{abstract}
\pacs{42.65.Re, 42.65.Tg, 42.50.Gy} 
\maketitle
The recent development of ultrafast science technology allowed the
generation of light pulses with durations down to few optical cycles. 
There is an increasing number of investigations and experiments which 
involve such ultrashort pulses in femtosecond and attosecond domains 
\cite{Brabec}. Applications also include control of chemical reactions 
in physicochemical processes \cite{phch}, femtochemistry \cite{fch}, 
biological multiphoton imaging \cite{bph} and coherent control schemes 
\cite{ccs}. In order to study light-matter interaction under extreme 
conditions as the pulse approaches the duration of a single optical cycle, 
great efforts for the generation of still shorter pulses have been made 
\cite{Leblond,Skobelev,Kalosha}. In particular, for continuous media the 
concept of single-cycle nonlinear optics \cite{scno} as well as the notion 
of extreme nonlinear optics \cite{eno} were introduced. For these processes, 
the traditional frame of the slowly varying envelope approximation (SVEA) 
and the rotating wave approximation (RWA) is invalid \cite{Rothenberg}. 

In the last decade, there has been a rapid progress in the field of 
materials with periodic structures, such as Bragg grating, photonic band 
gap crystals and waveguide arrays \cite{Busch}. Such systems exhibit
qualitatively novel and fascinating linear-optical, nonlinear-optical 
and quantum-optical properties which provides an attractive way to 
control the light propagation and light-matter interaction. Nonlinear 
pulse propagation in periodic structures is of both fundamental and 
applied interest, particularly, as a potential basis for nonlinear 
filtering, switching and distributed-feedback amplification \cite{Scalora}. 
Few-cycle pulse propagation in continuous media can yield solitons. 
Solitons can also exist in periodic structures which is usually referred 
to as gap soliton since they only exist within the forbidden gap 
\cite{Eggleton,Kurizki}. The gap soliton was primarily found in periodic
Kerr-nonlinear media. An essentially different mechanism of gap soliton 
generation with self-induced transparency has been revealed in Bragg 
grating which consists of a periodic array of thin layers of a 
resonant two-level medium separated by half-wavelength nonabsorbing 
dielectric layers \cite{Kozhekin,Mantsyzov}. It should be pointed out that 
for such a structure, the thickness of each two-level medium layer is much 
smaller than the wavelength of the electromagnetic wave which propagates. 
Optical soliton propagation results from an interplay between dispersion 
and the nonlinear response of the medium which balance the nonlinear and 
linear effects \cite{gsa}. Optical temporal solitons have become a promising 
candidate for optical communication networks. Thus, it is an important issue 
whether or not optical solitons can exist under extreme conditions, such as 
pulses with few optical cycles or less formed in a periodic structure. The 
ability to reduce the cycle number in these systems would be of particular 
significance.
\begin{figure}[b]
\includegraphics[width=8cm]{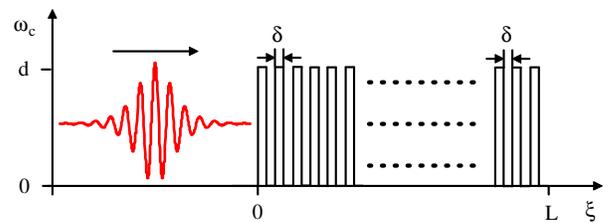}
\caption{\label{fig-1}(color online) Spatial distribution of atomic
medium. The red curve illustrates the incident wave with few optical 
cycles. Symbol "$\rightarrow$" denotes the propagation direction axis.}
\end{figure}

Therefore, in this Letter we present theoretical evidence of a sub-cycle 
gap soliton. An extremely short pulse propagating through a
one-dimensional periodic structure that consists of periodic thin
layers containing resonant two-level atoms separated by a
transparent material is considered (see Fig.~\ref{fig-1}).  
We assumed that the extremely short pulse propagates along the 
$\xi$ axis to an input interface of the periodic structure at 
$\xi = 0$ (a set of dimensionless time and space coordinates were 
taken, namely $T=t\omega_0$ and $\xi=z/\lambda_0$ with the carrier 
frequency $\omega_0$ and wavelength $\lambda_0$). First, the incident 
pulse moves in the free space and then partially penetrates into the
subwavelength structure. In the following, the penetrating part
passes through and exits into the free space. The length of the
structure is $L$. We found that at resonance and for a periodic 
structure satisfying the Bragg condition for moderate densities, 
the penetrating and reflected pulse spectrum is located around 
the transition frequency which is distinct from the results of 
few-cycle propagation in a continuous media. There, a large 
frequency shift of the transmitted and reflected pulses was 
reported \cite{Kalosha}. For particular parameters, we obtain a 
gap soliton with less than one optical cycle even when the 
structure period does not fulfill the Bragg condition. Thus, 
the system we propose allows the observation of different 
tendencies in the spectral and temporal evolution of ultrashort 
pulses leading to intresting and practically significant regimes 
of ultrashort-pulse quantum dynamics. Finally, these periodic structures 
can be realized by using a periodic doped fiber waveguide \cite{Kurizki} 
or quantum wells embedded in a semiconductor structure \cite{Watanabe}.

In order to investigate the electromagnetic field propagation in the 
periodic structure, we adopt the full Maxwell-Bloch equations solved 
by a finite difference time-domain method \cite{Ziolkowski} which has 
been proved to be an accurate tool. In this case, the evolution of 
the few-cycle electromagnetic field amplitudes $E_x(T,\xi)$ and $H_y(T,\xi)$,
the complex atomic polarization $\frac{1}{2}u(T,\xi)+\frac{i}{2}v(T,\xi)$, 
and the inversion population $w(T,\xi)$ obey the following set of 
Maxwell-Bloch equations: 
\begin{align}
\frac{\partial}{\partial T} H &=-\frac{1}{2\pi} \frac{\partial}{\partial \xi} E~, \nonumber\\
\frac{\partial}{\partial T} E &=-\frac{1}{2\pi} \frac{\partial}{\partial \xi} H - 
\frac{\omega_c(\xi)}{\Omega_0}\frac{\partial}{\partial T}u~,\nonumber\\
\frac{\partial}{\partial T} u &=-\frac{1}{\tilde T_{2}\omega_{0}}u -v~,\nonumber \\
\frac{\partial}{\partial T} v &=-\frac{1}{\tilde T_{2}\omega_{0}}v + u + \frac{2\Omega_0}{\omega_0}E~w~,\nonumber\\
\frac{\partial}{\partial T} w &=-\frac{1}{\tilde T_{1}\omega_{0}}(w + 1) -
\frac{2\Omega_0}{\omega_0}E~v~. \label{mb}
\end{align}
Here, $E=E_x/E_0$ and $H=\sqrt{\mu_0/\epsilon_0}H_y/E_0$ are the
normalized dimensionless components of the few-cycle pulse ($E_0$
corresponds to the peak amplitude of the electric component of the
incident pulse). $\Omega_0=gE_0/\hbar$ with the dipole moment $g$ 
is the peak Rabi frequency of the incident pulse. $\tilde T_{1}$ 
and $\tilde T_{2}$ are, respectively, the lifetime of the excited 
state and the dephasing time. In the above equations, we have 
assumed that the transition frequency of the two-level atom is equal 
to the carrier frequency $\omega_{0}$ of the pulse. Further, 
$\omega_c(\xi)=N(\xi)g^2/\epsilon_0\hbar$ with $N(\xi)$ denoting the 
density distribution of the atomic medium and which is given by the 
expression
\begin{eqnarray} 
\omega_c(\xi) = \left\{ \begin{aligned}
         d ~\mathrm{fs^{-1}} ~~&\mathrm{if}~~\xi\in[2n\delta,~~2n\delta+\delta ), \\ \label{ad}
                  0~~~~~~ &
                  \mathrm{if}~~\xi\in[2n\delta+\delta,~~2n\delta+2\delta), 
                          \end{aligned} \right. 
\end{eqnarray}
where $d$ and $\delta$ express the atom density and the thickness of
one layer. $n$ is an integer between $[0,~~L/(2\delta)-1]$.  
The density distribution of the atomic medium is shown in Fig.~\ref{fig-1}. 
The incident pulse we consider is a few-cycle pulse with a hyperbolic 
secant envelope, i.e.,
$\Omega(T=0,\xi)=\Omega_0\cos[2\pi(\xi-\xi_0)]\mathrm{sech}[1.76\times2\pi(\xi-\xi_0)/\tau_p]$,
where $\tau_p$ is the full width half maximum (FWHM) of the pulse
intensity envelope (see the red curve in Fig.~\ref{fig-1}). 
The choice of $\xi_0$ ensures that the pulse
locates in the free space and the partial pulse in the structure can
be neglected at $t=0$. The two-level medium is initialized with
$u=v=0$ and $w=-1$ at $t=0$. The material and laser parameters are
chosen as follows: $\tau_p=5~\mathrm{fs}$, $\omega_0=2.3~\mathrm{fs^{-1}}$ 
($\lambda_0=830~\mathrm{nm}$), $g=2\times10^{-29}\mathrm{A~s~m}$, 
$\tilde T_{1}=1~\mathrm{ps}$, $\tilde T_{2}=0.5~\mathrm{ps}$, and 
$N=4.4\times 10^{20}~\mathrm{cm^{-3}}$. The above parameters give an 
atomic density $d=0.2$. The peak Rabi frequency $\Omega_0=1.4~\mathrm{fs^{-1}}$ 
corresponds to an input envelope area $A=\Omega_0\tau_p\pi/1.76=4\pi$. The 
cycle number of the initial pulse is about $1.83$.
\begin{figure}[t]
\includegraphics[width=7.8cm]{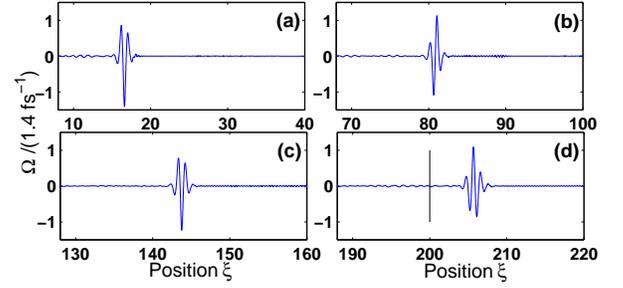}
\caption{\label{fig-2}(color online) Electric field profiles for
different time moments. (a) $t=0.1~\mathrm{ps}$; (b)
$t=0.3~\mathrm{ps}$; (c) $t=0.5~\mathrm{ps}$; (d)
$t=0.7~\mathrm{ps}$. The other atomic and structure parameters are: 
$\Omega_{0} = 1.4~\mathrm{fs^{-1}}$, $d=0.2$, $\delta = 0.25$ and 
$\xi \in [0,200]$. (a) and (d) correspond to the electric field 
profiles near the input face and outside the exit face (vertical 
line in d), respectively.}
\end{figure}

\begin{figure}[t]
\includegraphics[width=7.5cm]{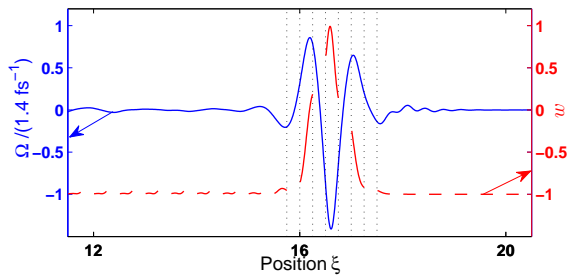}
\caption{\label{fig-3}(color online) Inversion population distribution 
(red curve) and electric field profile (blue curve) at time $t=0.1$ps. 
$\Omega_{0}=1.4~\mathrm{fs^{-1}}$, $\delta=0.25$, $d=0.2$ and $\xi \in [0,108]$.}
\end{figure}

We have obtained numerical results of Eqs.~(\ref{mb}) for different
periodic structures. Firstly, let us investigate the few-cycle
pulse propagation in a resonant structure in which the structure
period $2\delta$ exactly satisfies the Bragg condition: $2\delta =
m/2$ (the dimensionless coordinate is taken and $m$ is an integer
here). In our case, the period of the structure is set to be equal to
$1/2$. The electric field profiles of the penetrating pulse inside 
and outside the periodic structure for different time moments are 
shown in Fig.~\ref{fig-2}. As can be observed, the pulse can maintain 
the same envelope to propagate which is a typical behavior of the gap 
solitons in periodic structures. Remarkably, the cycle number of the 
transmitted pulse is reduced in comparison to the initial pulse. We 
obtain a $0.85$ cycle for the output pulse in Fig.~\ref{fig-2}(d). To 
further describe the system, we plot in Fig.~\ref{fig-3} the inversion 
as well as the electric field profile at particular time-space 
coordinates. The distribution of inversion $w$ is not continuous and 
in certain layers one can create population inversion. Therefore, the 
sideband parts of the propagating pulse annihilate while the central 
part slightly amplifies. Evidently, the output pulse energy is not 
increased in comparison to that of the initial applied pulse.

\begin{figure}[b]
\includegraphics[width=8cm]{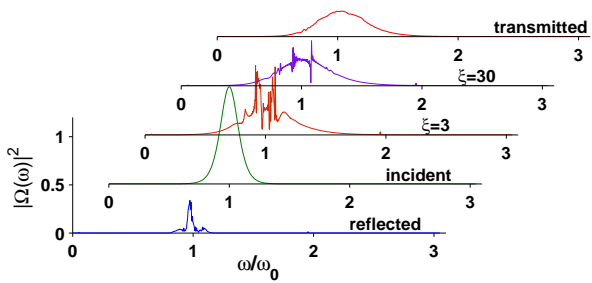}
\caption{\label{fig-4}(color online) Evolution of the pulse spectrum
during propagation inside and outside the periodic structure. The
atomic and structure parameters are the same as in Fig.~(\ref{fig-3}).}
\end{figure}

Subsequently, in order to clarify the soliton propagation in detail,
we present in Fig.~\ref{fig-4} the evolution of the pulse spectrum
during propagation through a periodic structure when $\delta=0.25$. 
Inspecting Fig.~\ref{fig-4}, we observe that the reflected and the
penetrating spectra are situated around the input central frequency 
$\omega_0$ which is distinct from the results of few-cycle propagation 
in the continuous media \cite{Kalosha}. There, for $d=0.2$ and $\xi=108$, 
the carrier frequency of the output pulse locates at $1.16\omega_{0}$. 
In our case, however, the large frequency shift of the transmitted and 
reflected pulses caused by intropulse four-wave mixing can be
efficiently suppressed. Periodic structures can usually offer optical 
band gaps similar to the flow of electrons in semiconductors and 
modulate the transmission of waves with different frequencies.
In the same way, the periodic structure in our model is able to
modify the dispersion property of the two-level atomic medium and
destroy the phase-matching condition of four-wave mixing process
which guarantees that the deformation of the penetrating pulse does
not happen in our model. From here, we can conjecture that the periodic
structure provides an alternative method for accurately preparing
few-cycle pulses which will allow one to approach the experimental 
study of single-cycle optics \cite{scno}. Thus, we have demonstrated 
that single sub-cycle gap solitons can be formed in a periodic structure.

\begin{figure}[t]
\includegraphics[width=7.5cm]{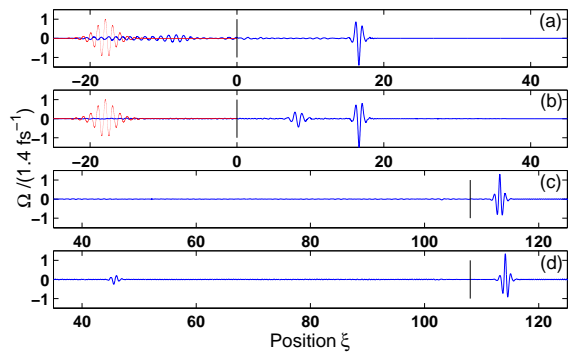}
\caption{\label{fig-5}(color online) Electric field profiles for
different time moments and structure periods near the input face
[(a),(b)] and outside the exit face [(c),(d)]. (a)
$t=0.1\mathrm{ps}$, $\delta=0.25$; (b) $t=0.1\mathrm{ps}$,
$\delta=0.1$; (c) $t=0.4\mathrm{ps}$, $\delta=0.25$; (d)
$t=0.4\mathrm{ps}$, $\delta=0.1$. Red line on the left of (a) and
(b): incident pulse at $t=0$. Here $\Omega_{0}=1.4~\mathrm{fs^{-1}}$, 
$d=0.2$ and $\xi \in [0,108]$.}
\end{figure}

In the following, we will discuss the case of a few-cycle pulse
propagation in a tighter periodic structure which violates the Bragg
condition. As a comparison, in Fig.~\ref{fig-5}, the initial,
reflected, penetrating and transmitted pulses near the input face
and outside the exit face are shown for different structure periods.
In the simulation, the parameters are the same except for the
structure period $2\delta$. For the case of tighter periodic
structure ($\delta=0.1$), the penetrating part is strong enough to
split into two pulses where the stronger of which moves faster than the
weaker one that is unstable and decays quickly (see Fig.~\ref{fig-5}b 
and Fig.~\ref{fig-5}d). More energy of the incident pulse is located in 
the structure in this situation which is different from the case 
of $\delta=0.25$ (compare Fig.~\ref{fig-5}a with Fig.~\ref{fig-5}b).
At the same time, by comparing with the profiles of the penetrating and 
transmitted pulses, one can observe that the incident pulse is dramatically 
shortened. We obtain the cycle number, $N_{c}$, for the FWHM of the intensity 
of the output pulse in Fig.~\ref{fig-5}(c,d) as $N_{c}\simeq 1.76\tau/(2\pi)$ 
with $\tau \approx 2.5$. Thus, the cycle number $N_{c}$ is about $0.7$ which 
indicates that the gap soliton with less than one optical cycle can be formed 
in a subwavelength structure. The pulse propagation for other cases 
(i.e. for longer pulses and structure periods, slightly different and random 
layer widths, etc.) has been simulated, too, and the behavior of the penetrating 
pulse is similar to the situations shown in Fig.~\ref{fig-2} - Fig.~\ref{fig-5}, 
though for longer pulses one can obtain fewer output stable pulses.
If the atomic density is increased, the single-cycle gap soliton cannot be 
obtained for larger structures. The reason for this is that the reflection 
on the interfaces of the atomic layers is enhanced rapidly with the increase 
in the atomic density. The output field spreads and becomes weak. Similar 
things can happen if one moderately increases the initial pulse intensity. 
However, shorter structures ($\xi \sim 20$) can lead to an output single-cycle 
pulse, too. If $0.2 < d < 1$, its carrier frequency slightly shifts towards 
higher frequencies when $\Omega_{0}$ approaches $\omega_{0}$ (the shift is 
smaller than that in a continuous sample with identical parameters).

In general, gap solitons are generated as a result of interaction
of counterpropagation waves. As remarked above, the existence of gap
soliton solution of the envelope equations is restricted by the
validity of SVEA and RWA which is violated for the conditions considered here. 
However, in the above two approximations, the pulse propagation in a resonant 
periodic structure can be described by a set of semiclassical coupled-mode 
Maxwell-Bloch equations. Then, assuming that the pulse consists of two strong 
coupled modes that propagate in the structure, one can obtain a {\it sech}-form 
soliton solution which has the following form \cite{Mantsyzov}:
$E^{\pm}=\hbar \Omega^{\pm}/g$, where 
\begin{eqnarray}
\Omega^{\pm}=\Omega^{\pm}_{m}\mathrm{sech}[(2\pi\xi-v_{g}T)/(v_{g}\tau)]
\cos(2\pi\xi\mp T+\phi). \label{RB}
\end{eqnarray}
Here, $E^{+}$ and $E^{-}$ denote the electric-field amplitudes of the
forward and backward waves. $v_{g}$ and $\tau$ are group velocity and
pulse duration, respectively. $\phi$ is an arbitrary phase factor while
$\Omega^{\pm}_{m}=\omega_{0}(1 \pm 1/v_{g})/\tau$. The relationship of 
parameters $v_{g}$ and $\tau$ is 
$v^{-2}_{g} = 1 + 2\omega_{c}\tau^{2}S/\omega_{0}$ with $S$ depending on
the broadened line form \cite{Mantsyzov}. With the help of these 
results, one obtains the envelope areas of the forward and backward 
waves: $A^{\pm}=\int\frac{\omega_0}{\tau}(1\pm v_g^{-1})\mathrm{sech}
[(2\pi\xi-v_gT)/(v_g\tau)] dt = \pi(1\pm v_g^{-1})$ which is referred 
to as gap $2\pi$ soliton \cite{Kozhekin,Mantsyzov}. If the field amplitude 
is strong enough, the cycle number of the stable soliton will be small or 
less than one.
\begin{figure}[t]
\includegraphics[width=7cm]{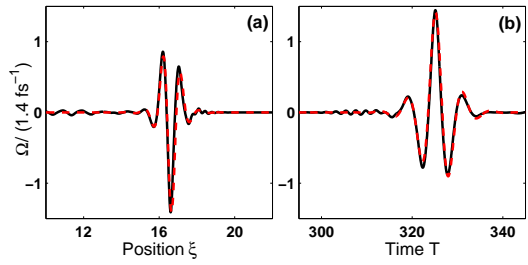}
\caption{\label{fig-6}(color online) Electric field profiles and
corresponding fitting curves in space and time coordinates (a) for
$t=0.1\mathrm{ps}$; (b) for $\xi=30$. The black and red lines
represent the numerical result and the analytic solution $\Omega^{+}$ 
given in Eq.~(\ref{RB}), respectively. $\Omega_{0}=1.4~\mathrm{fs^{-1}}$, 
$\delta=0.25$, $d=0.2$ and $\xi \in [0,~200]$.}
\end{figure}

As an example, in Fig.~\ref{fig-6}, the electric field profiles and
corresponding fitting curves in space and time coordinates are
illustrated. Here, the group velocity $v_{g}=(\xi_{2}-\xi_{1})/(t_{2}-t_{1}) 
\approx 0.85$ where $\xi_{2}$ and $\xi_{1}$ are the positions of the pulse 
maxima at $t_{2}$ and $t_{1}$, respectively, (see, for example, 
Fig.~\ref{fig-2}(a,c) whereas velocity unit is the light velocity in 
vacuum). The pulse duration is $\tau \approx 2.5$. Due to the fact that the 
pulse curves are mainly determined by the forward wave in the early stage of 
the formation of gap soliton, the fitting function is taken to be of the form 
of a forward wave $\Omega^{+}$, i.e., $\Omega^{+}=(\omega_{0}/\tau)(1 + 1/v_{g})\mathrm{sech}[2\pi(\xi-\xi_{m})/(v_g\tau)]\cos{\bigl(2\pi(\xi-\xi_{m})+\phi_{\xi}\bigr)}$ 
when $t$ is fixed while $\Omega^{+}=(\omega_{0}/\tau)(1+1/v_{g})\mathrm{sech}[-(T-T_{m})/\tau]
\cos{\bigl(-(T-T_{m})+\phi_{T}\bigr)}$ when $\xi$ is fixed. $\{\xi_{m},T_{m}\}$ 
and $\{\phi_{\xi},\phi_{T}\}$ describe the corresponding positions of the maxima 
of the profile pulse given by Eqs.~(\ref{mb}). In particular, we obtain 
$\xi_{m} \approx 16.6$, $T_{m} \approx 325.5$, $\phi_{\xi} \approx 0.87\pi$ and 
$\phi_{T} \approx 1.9\pi$ for the curves shown in Fig.~{\ref{fig-6}}. Here, one 
interesting result is that the {\it sech}-form solution derived from the 
semiclassical coupled-mode Maxwell-Bloch equations within the frame of SVEA and 
RWA coincides well with the numerical results in our model. With the further 
evolution of the penetrating pulse, the electric field profiles will depend on 
the forward and backward waves. 

In summary, we demonstrated the formation of a single sub-cycle gap
soliton when a few-cycle laser pulse penetrates a subwavelength 
structure. The carrier frequency of the output pulse is located around 
the frequency of the initial laser pulse. The proposed scheme works even 
for the case of broken Bragg condition as well as slightly different random 
layer widths. The corresponding analytical solution of the single-cycle 
gap soliton was given.

We acknowledge helpful discussions with Christoph H. Keitel and Karen 
Z. Hatsagortsyan.

\end{document}